\documentstyle{article}
\begin{document}
\baselineskip 14pt
\thispagestyle{empty}
{\bf \Large The Atomic and Electronic Structure of
Liquid N- Methyl\-form\-amide as Determined from
Diffraction Experiments}\\[1cm]

\noindent
{\large J. Neuefeind \footnote{ present address:
Hamburger Synchrotronstrahlungslabor HASYLAB at
DESY, Notkestrasse 85, 22603 Hamburg, Germany }
 and M. D. Zeidler}\\
 Institut f\"{u}r Physikalische Chemie,
RWTH Aachen, Templergraben 59, 52056 Aachen, Germany

\noindent
{\large H. F. Poulsen \footnote{ present address:
 Materials Dept., Ris\o\
National Laboratory, 4000 Roskilde, Denmark}} \\
Hamburger Synchrotronstrahlungslabor HASYLAB at
DESY, Notkestrasse 85, 22603 Hamburg, Germany

\vspace{1cm}

\vspace{1cm}
\begin{abstract}

The structure of liquid N-methylformamide (NMF) has been investigated
using synchrotron radiation at 77 and 95 keV.
The use of high energy photons has
several advantages, in this case especially the large accessible momentum
transfer range, the low absorption and the direct comparability with
neutron diffraction.
The range of momentum transfer covered is 0.6 \AA$^{-1}
<$ Q $<$24.0 \AA$^{-1}$.
Neutron diffraction data on the same sample in the same momentum
transfer range have been published previously.
In that study two differently isotope - substituted species were
investigated. In order to compare neutron and photon diffraction
data properly Reverse Monte Carlo (RMC-) simulations have been performed.
Some modifications had to be added to the standard RMC- code
introducing different constraints for inter- and intramolecular distances
as these distances partly overlap in liquid NMF.
RMC- simulations
having only the neutron data as input were carried out
in order to test the quality of the X-ray data. The photon structure
factor calculated from the RMC- configurations is found to agree well
with the present experimental data, while it deviates considerably from
earlier X-ray work using low energy photons (17 keV).
Finally we discuss whether the different interaction
mechanisms of neutrons and photons can be used to directly access the
electronic structure in the liquid. Evidence is presented that
the elastic self scattering part of liquid NMF is
changed with respect to the independent atom approximation.
This modification can be accounted for by a simple charged
atoms model.

\end{abstract}
\vspace{1cm}

\section{Introduction}
Liquid N-methylformamide (NMF) OHC'--NH(CH$_3$) is the simplest molecule
containing the peptide bonding system O=C'--N- relevant for the structure
of proteins. It has a very high dielectric constant (167.1 at
35 $^{\rm o}$C as compared to 74.8 for water). This fact has been
attributed to the ability
of forming hydrogen bonded chains aligning the molecular dipoles since the
early measurements of Cole {\it et al.} 1964 \cite{cole}.
The peptide system is believed to be essentially flat,
having a barrier to internal rotation around the N--C' bond due to
the delocalization of the O=C--N- $\pi$- bonding system \cite{pauling}.
The barrier to internal rotation gives rise to two different
isomers: cis- and trans- NMF (trans is defined to be the isomer
with the oxygen and the amide hydrogen on different sides
of the N--C'- bond). The trans-isomer is the more stable one being
present at room temperature to about 90-95\% in the liquid.

These interesting properties have lead in the past
to two liquid structure determinations -- in spite of the
relative complexity of the molecule giving rise to at least 28 independent
partial structure factors, when treating the methyl protons as equivalent.
First Ohtaki
{\it et al.} \cite{ohro} used conventional X-rays (Mo-K$_\alpha$: 17.4 keV)
for the structure determination and interpreted their data in
terms of chains of trans-NMF molecules. Next, neutron diffraction
was done on two different isotope-substituted species,
one being fully deuterated (NMF-d5), the other containing
one hydrogen nucleus in the amide position: ODC--NH(CD$_3$) (NMF-d4)
\cite{ego}(this reference henceforth referred to as I). The neutron data
will be used in the present paper to  check the quality of the
high energy photon data as well as an additional source of information
for the RMC simulations.

Surprisingly, the crystal structure of NMF has been determined only
recently \cite{vielraa}, whereas there is a couple of crystal structure
determinations available for similar molecules such as formamide
 \cite{post,ottersen,stevens}, acetamide  \cite{ottersen,hamilton,senti}
and N-methylacetamide \cite{katz}.
In the crystal NMF forms helical chains
of hydrogen bonded trans-molecules. The molecular structure in the gas phase
on the other hand has been determined by electron diffraction
\cite{kitkut}, microwave spectroscopy \cite{elzaro} and several
quantum mechanical calculations \cite{ottersen,suga,peri,lifson,fogarasi}.
As a result the C'=O bond is found to be substantially
shorter in the gas phase then in the solid (and also shorter than in the
liquid, as stated in I) indicating
a higher double bond character, whereas the C'--N bond is longer.

The interpretation of X-ray diffraction experiments on liquids is
typically done in the independent atom approximation neglecting
the redistribution of electrons on molecule formation. This fact
has been used by 'neutron diffractionists' to argue
that photon diffraction has a lower inherent precision than neutron
diffraction. Thinking positive one can hope to learn something
about the electronic structure in the liquid by comparing neutrons and
photons. However,
the effect of redistribution of the valence electrons is very difficult
to observe in a liquid
as it is small. Its observation therefore requires a high precision on
both the neutron and the photon side of the experiment. A
further difficulty is the fact, that  the absorption
cross sections for conventional
X-rays (8 to about 20 keV) in general are much larger
than cross sections for neutrons, so that photon diffraction has to be done
under conditions quite different from
 a typical neutron diffraction experiment
(reflection geometry or very thin samples instead of the large samples
used in neutron diffraction). As the photoelectric absorption
cross section approximately decreases with $E^3$ this difficulty can be
circumvented
by using high energy photons ($\sim$ 100 keV). The conditions
for performing such experiments in a reasonable
time have been established recently
\cite{ongerus}. NMF is a very promising object for observing
deviations from the independent atom approximations since 75\% of
its electrons are valence electrons and the polarization of the
bonds is expected to be high.

\section{Experimental Set-up}
The experiment was performed on the Hard X-ray triple axis diffractometer
recently build at HASYLAB \cite{roland1,roland2}.
The diffractometer was provisionally
installed on beamline BW7 at DORIS-III, a conventional wiggler beamline not
optimized for high energy use.  At the time of the experiment
electrons of 4.45 GeV were stored in DORIS-III.
Slightly different set-ups were used
for the low- and high-Q part of the spectra, one covering the range from
0.6-14.6 \AA$^{-1}$  at an incident photon energy
of 76.8(8) keV, the other from 8-24 \AA$^{-1}$ at 95.5(9) keV.
The energies were determined by calibrations of a multi channel
analyser with a $^{133}$Ba radioactive source and from the powder
lines of an (external) aluminium standard.
The small scattering angles at high energies allow
for a tangential move of the cooled Ge solid state detector.
No provision was made to exclude the air. The beamline contains no focussing
parts. The spectrometer was operated in a two- axis mode and no separation
of inelastic (Compton-) scattering was attempted.
The data acquisition was carried out in several scans in order
to minimize errors caused by beam fluctuations and drift in general.
The liquid NMF - vacuum distilled shortly before
the experiment - was contained in a cylindrical
glass tube of 3 mm diameter and 10 $\mu$m wall thickness.
The present set-up leads to count-rates from 10000-500 s$^{-1}$.
100000- 400000 counts were accumulated at 670 points in
about 20 h.
In general, the conditions were
quite similar to the previously described experiment on glassy SiO$_2$
\cite{ongerus}.

\section {Data correction and error estimation}

The raw data - after summing up the individual scans - are shown in figure
\ref{fig-raw}. These data were subsequently corrected for background
(mainly scattering from air and sample container), absorption,
multiple scattering, polarization of the incoming beam,
the variation of the solid angle seen by the detector as a result of the
changing sample to detector distance (tangential movement, see above) and
the increased detection efficiency for inelastically scattered
photons.
$\mu r$ is 0.027 at 76.8 keV (with $\mu$ being the total absorption
 coefficient and
$r$ the radius of the sample) and hence both absorption and
multiple scattering corrections are very small. Air scattering is
the dominant background contribution, especially at low Q's.
The main
error sources in this study are therefore believed to be the instabilities
induced by a different response of the main detector and the monitoring
system to the synchrotron beam fluctuations, the knowledge of the
degree of linear polarization
of the monochromatic beam (assumed to be known within 1\%), the
distance sample to detector (equally known approx. within 1\%) and
the energy calibration.
The mean energy shift of the inelastically scattered photons at
Q=23 \AA$^{-1}$ is 1.6 keV at 95.5 keV. The effect of an increasing
detection efficiency for those photons is estimated to be a 1\% effect.
For a more detailed description of the
data correction procedure the reader is referred to \cite{ongerus}.
The fully corrected and normalized intensity is presented in figure
\ref{fig-korrekt}.

\section{Simulation details}
In RMC- simulations \cite{MCG} a suitable number of atomic or molecular
entities is placed in a cubic (or differently shaped) simulation box.
A Markov- chain is generated  accepting
moves provided the move is improving the agreement with experiment, or
if a random number in the range $[0:1]$ is less
than $\sum_i^{N_{exp}} \exp (- \Delta \chi_i^2/2 \sigma_i^2)$
where the sum extends over all $N_{exp}$ independent experiments.
Otherwise the move is rejected.
$\Delta\chi^2$ is the difference between the $\chi^2$ of
the old and the new configuration, where $\chi^2_i$ is defined here as:
\begin{equation}
\chi^2_i = \sum^{N_{\rm points}}[ S(Q)_{conf,i} - S(Q)_{exp,i}]^2
\label{eq-chi2}
\end{equation}
where the sum extends over all $N_{\rm points}$ points in Q-space
, S(Q) is
defined by Eqs. \ref{eq-gsq} and \ref{eq-gsqx},
$\sigma^2_i$ is a measure of the assumed error in data set $i$,
the subscript $conf$ refers to the $S(Q)$ calculated from the
configuration generated and the subscript $exp$ to the experiment.
RMC simulations in polyatomic molecular systems are not as straightforward
as in atomic systems. It is accepted, that the molecules should
not be treated as rigid because the molecular part
of the scattered intensity is then reproduced poorly. If on the other hand
the atoms
are treated as free one has to introduce the
information that the atoms are connected and form a molecule.
The approach used by Radnai {\it et al.} \cite{radnai} based
entirely on coordination constraints was found to be unsuitable
in our case, since in NMF inter- and intramolecular distances
overlap. As a consequence two problems arise:
$a_1$ and $a_2$ are the distances between which a next neighbour
of type B is expected around an atom of type A {\it within} the molecule
( so that $a_1 \le r_{AB}({\rm intramolecular}) \le a_2$) and $a_3$ is
the smallest distance at which a molecule of type B on a {\it different}
molecule is expected around A. In the standard RMC code no distinction
is made between inter- and intramolecular distances, thus even if
$a_1 \le a_2 \le a_3$ error messages occur as the intramolecular distances
between $a_1$ and $a_2$ violate the constraint that no AB distance should
be present below $a_3$ (hard core radius). This is still not disadvantageous,
but if $a_1 \le a_3 \le a_2$ the RMC code tries to satisfy both the
coordination constraint between $a_1$ and $a_2$ and the hard core radius
$a_3$. As a result all intramolecular distances are now comprised
between $a_3$ and $a_2$ what is not intended. If finally $a_3 \le a_1 \le
a_2$ there are real {\it inter}\/molecular distances between $a_1$ and
$a_2$ and hence the total coordination in the range $[a_1:a_2]$
is no longer known.
The only way to treat this problem is to fix
explicitly which atoms belong to the same molecule and to
give different coordination constraints and cut-off radii
for inter- and intramolecular distances. This can not be
done with the standard RMC, version III code\cite {RMCA}.

In the present simulation a 9 site model with 216 molecules corresponding
to 1744 atoms was used. The distance of closest approach was
fixed to 2.0 \AA\ for all intermolecular distances except
the O$\cdots$H--N were it was 1.5 \AA. The molecular density
was taken to be 0.01022 \AA$^{-3}$ corresponding to a half box  length
of 13.85 \AA. The intramolecular constraints are given in Table
\ref{tab-constraints} and are chosen to give  essentially planar
trans-molecules, where the methyl protons are free to rotate
around the N-C bond. The range in which the
intramolecular distances can fluctuate is chosen to be
four times the Debye-Waller factor as given by Kitano {\it et al.}
\cite{kitkut}.

\section{Analysis}
In order to assess the quality of the high energy photon data a RMC
simulation (modified as described above)
has been carried out using only the neutron data from I
as input. From the configurations generated we calculated the
corresponding photon scattering. We compare the
result to our  experimental data as well as to the data of Ohtaki
{\it et al. } \cite{ohro} obtained with Mo-K$_\alpha$ X-rays.

The  RMC fit to the two neutron data sets and its first
order difference is shown in figure \ref{fig-RMC}
in Q-space. The
definitions of the functions shown are:
\begin{eqnarray}
\label{eq-gsq}
S(Q)&=&\frac{\left ( \displaystyle \frac{d\sigma}{d\Omega} \right )_{dist}}
{(\sum_{N_M} b_i)^2}\\ \nonumber
g_n(r)&=&1+\frac{1}{2\pi^2\rho r} \int_0^{Q_{max}} Q S(Q) \sin(Qr) dQ,
\end{eqnarray}
where $(d \sigma /d \Omega)_{dist}$ is the distinct differential cross
 section, the sum is extended to all $N_M$ atoms in the molecule,
$b_i$ is the neutron scattering length of atom $i$, $\rho$ is the
macroscopic number density, $Q_{max}$ is the largest accessible momentum
transfer where $S(Q)=0$, $S(Q)$ is the neutron weighted structure
factor and $g_n(r)$ is the neutron weighted pair distribution function.
 The agreement between simulation and experiment
is good within the statistics given by the number of sites and the
chosen $\sigma$- parameter for the total structure factors, but not
perfect for the hydrogen first order difference in the low-Q region.
The first order difference is given here by:
\begin{equation}
S_{N-H} = S_{d5} - \frac {(\sum_{N_M(d4)}b_i)^2}
{ (\sum_{N_M(d5)}b_{i})^2} S_{d4}
\end{equation}
Note, however, that the first order difference has been enlarged
 by a factor of 5. The fit is considered to be good enough for
our purposes ($\sqrt{\chi^2 /N_{\rm points}}
$ defined by Eq. \ref{eq-chi2}
is 0.0032 for NMF-d5 and 0.0043 for NMF-d4).

The next question to ask is whether or not
the configurations consistent with the neutron data are also consistent
with the photon diffraction data. With definitions  analogous to those
appearing in Eq. 2:
\begin{eqnarray}
\label{eq-gsqx}
i(Q)&=&\frac{\left ( \displaystyle \frac{d\sigma}{d\Omega} \right )_{dist}}
{(\sum_{N_M} f_i)^2}\\ \nonumber
g_x(r)&=&1+\frac{1}{2\pi^2\rho r} \int_0^{Q_{max}} Q i(Q) \sin(Qr) dQ
\end{eqnarray}
the comparison RMC(neutrons) {\it versus} experiment(photons) is done in Fig.
\ref{fig-RMCexp1} and \ref{fig-RMCexp2}. $i(Q)$ is here the
photon weighted structure factor, $g_x(r)$ the photon weighted
pair distribution function and $f_i$ the form factor of the $i^{\rm th}$
atom.

Focussing first on the real space comparison two things should be noted:
First, $g_x(r)$ is not equal to zero at $r \le r_{cut}$  for the
Fourier- transform of the RMC $S(Q)$ - where $r_{cut}$ is
the smallest internuclear distance. This is known to be caused
by the fact that the terms $f_if_j/(\sum f)^2$ are not constant in
$Q$. Second,
there is a much larger deviation from zero for the Fourier-transform
of the experimental $S(Q)$. It is current practice to  back
 Fourier-transform
the deviation of $g_x(r)$ from zero at low $r$ to get an idea
about the size and the location of
systematic errors in Q-space \cite{xpapst}
However, this approach is known to be incorrect, as $g_x(r)$
may deviate from zero even for data completely free from systematic errors.
We have therefore chosen to use the following 'correction'-function instead:
\begin{equation}
K(Q)=(\sum f)^2 \frac{4 \pi \rho}{Q}
\int_0^{r_{cut}} [g_x(exp)-g_x({\rm RMC})] \sin (Qr) dQ
\label{gl-korr}
\end{equation}
$K(Q)$ will be shown to be largely related
to deviations from the independent atom approximation.

The form factor of a spherically symmetric atom is given by \cite{unkn}:
\begin{equation}
\label{elde}
f_{iaa}(Q)= \sum_{i=1}^Z 4\pi/Q \int_0^\infty  \psi_i^2(r) r \sin(Qr) dr
\end{equation}
where $f_{IAA}$ is the form- factor of the unperturbed atom, $Z$ is the number
of electrons within that atom and $\psi_i$
is the wave-function of the i$^{\rm th}$
electron. Using Roothaan-Hartree-Fock (RHF) atomic wave-functions as
tabulated by
Clementi {\it et al.} \cite{clementi}
the above integral can be solved analytically.

Now leaving the independent atom approximation,
a very simple approximation of atoms in molecules is the assumption
that the electron cloud of the atoms keep  its form
and spherical symmetry
but that the atoms can adopt net charges \cite{elde}. As a consequence
the valence shell of the respective atom is contracted or
expanded, while the core electrons remain unaffected:
\begin{equation}
\label{eq-char}
\psi'^2_{valence}(r)=\psi^2_{valence}(\kappa r)
\end{equation}
The simplicity of this approach keeps the integrals of Eq. \ref{elde}
to be solved analytically. The calculations are presented in the appendix.
 A comparison between
$(\sum f_i)_{iaa}^2-(\sum f_i)_{cm}^2$ and $K(Q)$
is done in Fig. \ref{fig-kf} (iaa for independent atom approximation
and cm for charged atoms model). The charges of
the atoms have been taken from formamide in the solid state
\cite{stevens} and it is assumed that the methyl group takes the charge
of one amide hydrogen. The charge parameters are given
in Table \ref{tab-charge}.
Using now the modified form-factors $f_{i,cm}$ in the analysis
the low $r$ deviations of $g_x(r)$ are very much reduced as
shown in Fig. \ref{fig-RMCexp2}, except for the peak at 0.3 \AA.
Given the crude model used, this result is very satisfying.
One can ask now whether the peak at 0.3 \AA\ also has a
physical meaning or is caused by experimental errors,
but further analysis seems to be speculative without supporting
quantum-mechanical calculations.

In Fig. \ref{fig-compohro} $i(Q)$ from our experiment and the one from
Ohtaki's experiment \cite{ohro} are compared to the RMC result
(with the neutron measurements as input only).
The discrepancies of Ohtaki's data in the first diffraction
peak are immediately evident. This is rather astonishing as in
the region $Q >3 {\rm \AA}^{-1}$ - dominated by intramolecular distances -
the agreement between Ohtaki's data and
the simulation is very good, even better than the agreement achieved
in the present study. It is noted that Ohtaki's data
do not extrapolate to the correct low Q limit given by the compressibility.
On the other hand the agreement between the high energy
photon result and the shape of the first maximum predicted by the
neutron result is nearly perfect. Concluding, there seems to be a larger
problem in Ohtaki's data analysis and the conclusions drawn in
his paper should therefore be accepted with care.

Including the hard X-ray data in the RMC simulations, does not
remove the slight discrepancies between simulation and experiment
in the range from 3-8 \AA$^{-1}$ still present in Fig. \ref{fig-compohro}. From
another point of view - these discrepancies are not of a
type which could be removed by simply moving atoms in the box
within the given constraints. Hence, as stated above, these discrepancies
are caused either by systematic errors still present or by deviations
form the independent atom approximation not accounted for by the
primitive model used. This fact also highlights, that the x-ray
diffraction pattern is {\it predicted} by the neutron diffraction data,
and the x-ray data are in excellent agreement with the neutron data.

Likewise, introduction of 10\% cis- molecules does not substantially
improve the fit. Nevertheless, the configurations generated in
this way might be closer to reality, as it is known
that a certain percentage of cis-molecules are
present in the liquid. The cis/trans isomerism only changes
rather long intramolecular distances - smeared by large amplitude vibrational
motions - which can be
modelled by intermolecular distances without coming into conflict
with the imposed constraints.

Finally, one aim of the present study was to test the conclusion drawn in I,
that the molecular structure of liquid NMF is more like in
the solid and not like the gas phase structure, hence,
answering the question, whether long range order is a necessary condition
for the found increased delocalization of the
$\pi$-electron bonding system. The $r$-space
resolution should be sufficient to
resolve the C'--N and the  C'=O distances if the gas- phase model of the
molecule is valid and at high Q the respective interference pattern
 should be observed. Different from the  neutron case these
distances are not significantly obscured by hydrogen-
heavy atom distances. This is illustrated in Fig. \ref{fig-gas}, but it is
also apparent from the figure that the counting statistics reached at high
Q-values so far is not sufficient to support the finding from I.  The
beam fluctuations mentioned in chapter 3 also contribute to the uncertainty in
S(Q). In this respect, the data quality will
probably improve in the near future,
especially with the introduction of an  area detector. It is also noted, that
the Q-range up to about 14 \AA$^{-1}$ covered by
conventional X-rays is not very conclusive for answering the question posed.

\section{Conclusion}

The structure of liquid NMF has been investigated using the
recently reported methods for high energy photon diffraction.
The structure factor derived is shown to be in much
better agreement with the information available from
neutron diffraction measurements than the earlier study
of Ohtaki {\it et al.} using low energy photons (conventional
Mo-K$_\alpha$-radiation). In the intermediate Q-range from
about 3-8 \AA$^{-1}$ some discrepancies between the experimental
self- scattering intensity and the self-scattering predicted
by the independent atom approximation are observed.
These discrepancies are accounted for by a simple charged atoms
model, based on charge parameters from solid state formamide.
The introduction of charged atoms largely improves the
low-r behavior of $g_x(r)$. Although promising for testing
a finding from neutron diffraction - that the intramolecular
structure of liquid NMF is closer to the solid state structure
than to the structure of the gas phase - this aim could not be reached
as the counting statistics were not good enough at high Q values.

High energy photon diffraction is likely to see
considerable progress and will be a useful tool for understanding
the physics of systems without long-range order. Its most
outstanding property will be the ease with which information from
neutron and photon diffraction can be combined.

\section{Appendix}
In this appendix the solution of the integrals appearing in Eq. \ref{elde}
will be sketched, first in the independent atom approximation using RHF
wave-functions, then in the charged atoms model given by Eq. \ref{eq-char}.
RHF wave-functions are of the form:
\begin{equation}
\psi_i=\sum_p C_p \chi_p
\end{equation}
where $C_p$ are the expansion coefficients of the basis functions $\chi$
which are of the Slater type:
\begin{equation}
\chi_{p\lambda\alpha}(r,\theta,\phi)=R_{p\lambda\alpha}(r)
Y_{\lambda\alpha}(\theta,\phi)
\end{equation}
with
\begin{equation}
R_p(r)=\mbox{const.}(\xi_p)\ r^{n-1} e^{-\xi_p r}
\end{equation}
where n is the quantum number of the basis function,
$\lambda$ indicates the symmetry species, $\alpha$ the subspecies
of the electron wave-function and $\xi_{p\lambda\alpha}$
the orbital exponent (The angular dependence of the 2p-type orbitals
averages out). Inserting the RHF wave-functions
into Eq. \ref{elde} creates integrals of the type:
\begin{equation}
\frac{1}{Q} \mbox{const.} \int_0^\infty r^{n_1+n_2-1}
e^{- (\xi_1+\xi_2) r} \sin(Qr) dr
\end{equation}
which can be solved analytically \cite{bronstein}.
With e. g. $n_1=1\ \mbox{and}\ n_2=2$
the above integral equals to:
\begin{equation}
\label{term}
\frac{8}{\sqrt{3}}\frac{C_1 C_2}{Q} \xi_1^{1.5} \xi_2^{2.5}
[(\xi_1+\xi_2)^2 + Q^2]^{-1.5} \sin (3 \arctan [Q/(\xi_1+\xi_2)])
\end{equation}
When modifying the wave-functions as in Eq. \ref{eq-char} the term
in Eq. \ref{term} e. g. has to be modified to:
\begin{equation}
\frac {8}{\sqrt{3}}\frac{C_1 C_2}{Q} \xi_1^{1.5} \xi_2^{2.5}
[(\xi_1+\xi_2)^2 + \left ( \frac{Q}{\kappa} \right ) ^2]^{-1.5}
\sin (3 \arctan [\frac{Q}{\kappa}/(\xi_1+\xi_2)])
/\kappa^2
\end{equation}
 It should be noted that the $Q\rightarrow 0$ limit of $\sum f^2$
in the charged atoms model is no longer $\sum Z_i^2$ but $\sum (Z_i+q_i)^2$,
where $Z_i$ is the nuclear charge and $q_i$ are the partial charges of
the atoms. $q_i$ is related to $\kappa$ by:
\begin{equation}
q_i= Z - [ N_{\rm valence}/\kappa^3 + N_{\rm core} ]
\end{equation}
where $N_{\rm valence}$ is the number of valence electrons
in the independent atom and $N_{\rm core}$ is the number of core electrons.

\section{Acknowlegement}

Good diffraction experiments are not only a question of photon
energy: The excellent working conditions at HASYLAB are gratefully
acknowledged. Thanks are especially due to J. R. Schneider and his group.

Financial support of the Fonds der Chemischen Industrie
is gratefully acknowledged.

\newpage

\begin{figure}
\caption{The high energy X-ray raw data}
\label{fig-raw}

 From top to bottom: scattering from NMF in a glass tube, from the
empty glass tube, and background
scattering (mainly air scattering). The high-Q data are divided by 10
for better visibility. The ratio counts/monitor counts is larger
for the high-Q scans as the distance sample/detector is shorter
and hence the sampled solid angle larger.
\end {figure}
\begin{figure}
\caption{The fully corrected and normalized data}
\label{fig-korrekt}

Full line: isotropic part of the scattering intensity (Self scattering
 + Compton part); broken line: experimental intensity. Note that the
slightly higher level of the isotropic part in the high-Q region
is caused by the higher energy (lower relativistic correction) of
the photons used. Insert: Zoom into the region Q $>$ 6 \AA$^{-1}$.
\end{figure}
\begin{figure}
\caption{RMC- fit to the neutron data}
\label{fig-RMC}

 From top to bottom: Fit to the NMF-d5 data, fit to the NMF-d4 data,
the resulting hydrogen first order difference. Full line: RMC simulation,
broken line: experimental data from I. The first order difference
is enhanced by a factor of 5.
\end{figure}
\begin{figure}
\caption{Comparison RMC(neutron data) {\it versus} experiment
(high energy photons)}
\label{fig-RMCexp1}

Full line: high energy photons, self scattering in the independent
atom approximation, broken line: high energy photons, self scattering
 from the charged atoms model (see text), points: RMC simulation, neutron
data as input only.
\end{figure}
\begin{figure}
\caption{Comparison RMC(neutron data) {\it versus} experiment
(high energy photons)}
\label{fig-RMCexp2}

Same as Fig. \ref{fig-RMCexp1}, but in real space.
\end{figure}
\begin{figure}

\caption{Comparison $K(Q)$ {\it versus}  $[(\sum f_{iaa})^2 - \sum(f_{cm})^2]$}
\label{fig-kf}

Full line: $K(Q)$ is a 'correction'- function
defined in the text, broken line: difference
between the self-scattering calculated from the independent atom approximation
 and the charged atoms model.

\end{figure}
\begin{figure}
\caption{Comparison of low energy photons with high energy photons and
neutrons}
\label{fig-compohro}

Full line (with diamonds): low energy photons (from Ohtaki {\it et al.}
\cite{ohro}, broken line (longer lines): RMC simulations, neutron
data as input only, broken line (shorter lines): high energy photons
(this work), self scattering in the charged atoms approximation,
dots: high energy photons (this work), self scattering in the independent
atom approximation.
\end {figure}
\begin{figure}
\caption{Comparison gas-phase model {\it versus}
 solid state model of the intramolecular structure}

Diamonds: experimental points, solid line: intramolecular i(Q) based
on the gas phase model with C'O = 1.22 \AA\ and C'N = 1.35 \AA, broken line:
intramolecular i(Q) based on the solid state model with C'O = 1.26 \AA\
and C'N = 1.29 \AA, points: smoothed experimental points.
\label{fig-gas}
\end{figure}

\begin{table}[htbp]
\caption{Constraints used for the RMC simulations}
\vspace{.7cm}

\begin{tabular}{|c|cc|cc|cc|cc|cc|cc|cc|}
\hline
&\multicolumn{2}{c|}{N}&\multicolumn {2} {c|} {N-$\underline{\rm H}$}&
\multicolumn{2}{c|}{C'}&\multicolumn{2}{c|}{C}&\multicolumn{2}{c|}{O}
&\multicolumn{2}{c|}{C'-$\underline{\rm H}$}&
\multicolumn{2}{c|}{C-$\underline {\rm H}$}\\ \hline
N-$\underline{\rm H}$&0.86&1.15&\multicolumn{2}{c|}{---}&1.78&
2.16&1.92&2.33&2.97&3.33&1.86&2.48&\multicolumn{2}{c|}{---}\\
C'&1.21&1.38&1.78&2.16&\multicolumn{2}{c|}{---}
&2.29&2.58&1.19&1.35&0.94&1.26&\multicolumn{2}{c|}{---}\\
C&1.36&1.56&1.92&2.33&2.29&2.58&\multicolumn{2}{c|}{---}
&2.64&3.09&3.10&3.51&0.91&1.23\\
O&2.17&2.38&2.97&3.33&1.19&1.35&2.64&3.09&\multicolumn{2}{c|}{---}
&1.89&2.25&\multicolumn{2}{c|}{---}\\
C'-$\underline{\rm H}$&1.79&2.18&1.86&2.48&0.94&1.26&3.10&
3.51&1.89&2.25&\multicolumn{2}{c|}{---}&\multicolumn{2}{c|}{---}\\
C-{$\underline{\rm H}$}&1.89&2.30&\multicolumn{2}{c|}{---}&
\multicolumn{2}{c|}{---}&
0.91&1.23&\multicolumn{2}{c|}{---}&\multicolumn{2}{c|}{---}&1.48&2.00\\
\hline
\end{tabular}
\label{tab-constraints}

\small All distances in \AA. Left number lower limit, right number
upper limit for the respective intramolecular distance.

\end{table}

\begin{table}
\caption{Charge parameters for the charged atoms model (cm)}
\vspace{.7cm}

\begin{tabular}{|c|ccccccc|}
\hline
&N&N-$\underline{\rm H}$&C'&C&O&C'-$\underline{\rm H}$&C-$\underline{\rm H}$\\
\hline
$\kappa$&0.9567&1.1856&1.0465&1.0093&0.9712&1.0102&1.0357\\
$q_i$&--0.78&+0.40&+0.51&+0.10&--0.55&+0.03&+0.10\\
\hline
\end{tabular}

\label{tab-charge}
\end{table}

\end {document}